\documentclass[%
 reprint,
superscriptaddress,
showpacs,preprintnumbers,showkeys
 amsmath,amssymb,
prx,
]{revtex4-1}

\usepackage{graphicx}
\usepackage{dcolumn}
\usepackage{bm}
\usepackage{sistyle}
\usepackage{color}
\SIthousandsep{,}

\renewcommand{\BibitemShut}[1]{}  
\begin{document}

\title[Bulk-mediated diffusion strange kinetics]{Strange kinetics of bulk-mediated diffusion on lipid bilayers}

\author{Diego Krapf}
\email[E-mail: ]{krapf@engr.colostate.edu}
\affiliation{Department of Electrical and Computer Engineering, Colorado State University, Fort Collins, CO 80523, USA}
\affiliation{School of Biomedical Engineering, Colorado State University, Fort Collins, CO 80523, USA}

\author{Grace Campagnola}
\affiliation{Department of Biochemistry and Molecular Biology, Colorado State University, Fort Collins, CO 80523, USA}

\author{Kanti Nepal}
\affiliation{School of Biomedical Engineering, Colorado State University, Fort Collins, CO 80523, USA}

\author{Olve B. Peersen}
\affiliation{Department of Biochemistry and Molecular Biology, Colorado State University, Fort Collins, CO 80523, USA}

\date{\today}
\begin{abstract}
Diffusion at solid-liquid interfaces is crucial in many technological and biophysical processes. 
Although its behavior seems deceivingly simple, recent studies showing passive superdiffusive transport 
suggest diffusion on surfaces may hide rich complexities. In particular, bulk-mediated diffusion occurs 
when molecules are transiently released from the surface to perform three-dimensional excursions into the 
liquid bulk. This phenomenon bears the dichotomy where a molecule always 
return to the surface but the mean jump length is infinite. Such behavior is associated with a breakdown of 
the central limit theorem and weak ergodicity breaking. Here, we use single-particle tracking to study the 
statistics of bulk-mediated diffusion on a supported lipid bilayer. We find that the time-averaged mean 
square displacement (MSD) of individual trajectories, the archetypal measure in diffusion processes, 
does not converge to the ensemble MSD but it remains a random variable, even in the long observation-time 
limit. The distribution of time averages is shown to agree with a L\'{e}vy flight model. 
Our results also unravel intriguing anomalies in the statistics of displacements. The time averaged MSD 
is shown to depend on experimental time and investigations of fractional moments 
show a scaling $\langle |r(t)|^q\rangle \sim t^{q\nu(q)}$ with non-linear exponents, 
i.e. $\nu(q)\neq\textrm{const}$. This type of behavior is termed strong anomalous diffusion and is 
rare among experimental observations.
\end{abstract}


\maketitle

\section{Introduction}

Processes at solid-liquid interfaces play important roles across multiple fields.
In particular surface diffusion and diffusion-controlled reactions have key functions 
in life sciences and biomedical technologies \cite{castner2002biomedical}. 
For example, surface reactions are of utmost 
importance in the development of implant biomaterials \cite{ratner2004biomaterials,hench2002third}, 
affinity chromatography methods \cite{Chaga}, 
and biosensors as well as in blood-contacting devices \cite{courtney1994} such as heart valves and 
hemodialysis membranes. In cell biology biomolecular recognition and reactions on surfaces are essential for 
a vast array of physiological functions. The importance of molecular films in biology 
has been discussed for more than a century \cite{loeb1922proteins}.
In fact, most biochemical reactions in cells take place 
at interfaces instead of in solution. Diffusion-controlled reactions often involve a search for a reactive target 
with the goal of minimizing the search time \cite{hanggi1990reaction,benichou2011intermittent}. 

The random motion of a particle is usually characterized by the 
mean squared displacement (MSD). In its simplest form, 
diffusion processes can be described by Brownian motion, which in two dimensions (2D) manifests a linear MSD 
$\langle r^2 (t)\rangle = 4Dt$, where $D$ is the diffusion coefficient. 
However, diffusion at solid-liquid interfaces can exhibit rich complexities 
\cite{weigel2011PNAS,barkai2012PhysToday,skaug2013intermittent,metzler2014review,krapf2015reviewCTM}.
Systems with a non-linear MSD $\langle r^2 (t)\rangle = K_{\alpha} t^{\alpha}$ 
display anomalous diffusion, where a slower-than-linear growth, i.e $\alpha<1$, 
indicates subdiffusion; and faster-than-linear growth, $\alpha>1$, indicates superdiffusion. 
Most importantly, anomalous diffusion alters reaction kinetics because the diffusion properties 
control the rate of molecular encounters \cite{shlesinger1993strange,benichou2010geometry}.

A widespread feature of molecules diffusing at the solid-liquid interface 
involves the desorption of molecules from the surface into the liquid phase. 
Molecules will diffuse in three dimensions (3D) until they reach the interface again and 
readsorb. This intermittent process where molecules alternate between 2D and 3D phases 
is known as bulk-mediated diffusion and has been previously analyzed in terms of scaling arguments \cite{Bychuk1995},
simulations \cite{bychuk1994anomalous,revelli2005bulk}, and analytical approaches \cite{chechkin2012}. 
Recently bulk-mediated diffusion was experimentally observed in systems of vastly different nature including 
organic molecules at chemically coated interfaces \cite{skaug2013intermittent,skaug2014single}, 
polymer-surface interactions \cite{yu2013single}, 
and membrane-targeting domains on both 
supported lipid bilayers \cite{knight2009BJ,campagnola2015superdiffusion} 
and the plasma membrane of living cells \cite{yasui2014pten}.
Diffusion as measured on the surface is strongly influenced by the statistics 
of excursion times. 
On each excursion a random distance is covered on the surface, which scales 
in probability as the square root of the return time ($\langle r^2 (t)\rangle = 4D_b t$).
The first return time to the surface has interesting properties \cite{redner}. 
The most fundamental of these properties is the dichotomy between mean first return time and probability 
of return. On one hand, the mean first return time is infinite due to its heavy tail distribution 
$p(t)\sim t^{-1.5}$. On the other hand, a particle always returns to the surface, that is the 
probability of return is one. In terms of probability theory one would say the particle returns 
to the surface almost surely. To place the problem in real context, if we consider  
a generic protein that alternates between a lipid bilayer and a water-based solution, the probability that 
it returns to the surface within less than 50 ms after it reached a 10-nm height 
is 99.75\% \cite{campagnola2015superdiffusion}.

A diffusion process where long jumps with a heavy-tail distribution occur is known 
as L\'{e}vy walk \cite{shlesinger1986levy}. 
In such a random walk, jumps are performed at a velocity that might depend on the jump distance 
\cite{shlesinger1987levy,klafter1996beyond}. 
If the long jumps take place instantaneously, the process is known as L\'{e}vy flight \cite{mandelbrot1983fractal}. 
L\'{e}vy walks have traditionally received more attention than flights because  
instantaneous jumps are not realistic. 
However, in the limit where bulk diffusion is orders of magnitude faster than surface diffusion, 
$D_b>>D_s$, a L\'{e}vy walk can be approximated as a L\'{e}vy flight, at least within short time scales. 
This regime is found to be the most relevant for experimental observations of bulk-mediated diffusion.  

Both L\'{e}vy flights and walks are superdiffusive when the probability density of jump distances 
scales as $p(r)\sim r^{-(1+\beta)}$ with $\beta \le 2$. 
We recently reported that the motion of membrane-targeting domains 
on lipid bilayers is superdiffusive due to bulk excursions \cite{campagnola2015superdiffusion}. 
In these experiments, the MSD grows faster-than-linear when 
it is measured over an ensemble of molecules, that is the average is performed by employing 
a single displacement for each trajectory at any given time. Nevertheless, when the average is performed 
over time, i.e. by averaging all the displacements observed along a trajectory, 
the MSD is linear in lag time. This observation contradicts the ergodic hypothesis, one of the 
cornerstones of statistical mechanics, which states that ensemble 
averages and long-time averages of individual trajectories are equivalent.
A similar behavior is found in subdiffusive continuous time random walks (CTRWs), 
where the ensemble-averaged MSD follows a power law $t^\alpha$, but the time-averaged
MSD is linear \cite{lubelski2008nonergodicity,he2008random}. 
In the CTRW, the non-ergodic property is rooted in 
the system not being stationary. Such strange kinetics where the random walk
exhibits different scaling properties depending on whether it is averaged over time or over 
an ensemble poses intriguing questions regarding its statistics. 
Beyond the MSD, the distribution of displacements also deviates from 
``normal'' diffusion. The central limit theorem (CLT) warrants that the displacements of Brownian motion
have a Gaussian distribution. 
However, in some types of anomalous diffusion models, the CLT breaks down 
and the distribution of displacements is no longer Gaussian. For example, in a 
CTRW or when a particle diffuses in a fractal structure, the increments are not independent 
and thus the CLT does not hold. In a L\'{e}vy flight the CLT breaks down because the 
increments can have infinite variance \cite{Bychuk1995,chechkin2012}.

Here we investigate the kinetics of membrane-targeting C2 domains on lipid bilayers using single-particle
tracking. This system exhibits superdiffusive behavior in the 
ensemble-averaged MSD but normal scaling in the time-averaged MSD. 
Weak ergodicity breaking predicts large fluctuations in the time-averaged MSD of individual 
trajectories. Thus we examine the fluctuations in the MSD and find that it remains 
a random variable even in the long time limit. In contrast to the CTRW model, the 
increments of bulk-mediated diffusion are shown to be stationary, but the statistics of the motion still depend on experimental 
time. It is found that when the MSD is averaged over both time and ensemble, it does not 
converge to a finite value, but it increases with experimental time. Thus,
if the diffusion coefficient were estimated using the MSD slope, it would increase as the experimental 
time increases. The experimental results for 
bulk-mediated diffusion are found to agree with a L\'{e}vy flight model using both analytical approaches and 
numerical simulations. Interestingly we also find the system exhibits 
strong anomalous diffusion \cite{castiglione1999strong}, i.e., the fractional moments 
are not characterized by a linear scaling exponent as in most diffusion processes.     

\section{Experimental Results}

\subsection*{Fluctuations in time-averaged MSD}
\begin{figure}
\centerline{\includegraphics[width=0.6\linewidth]{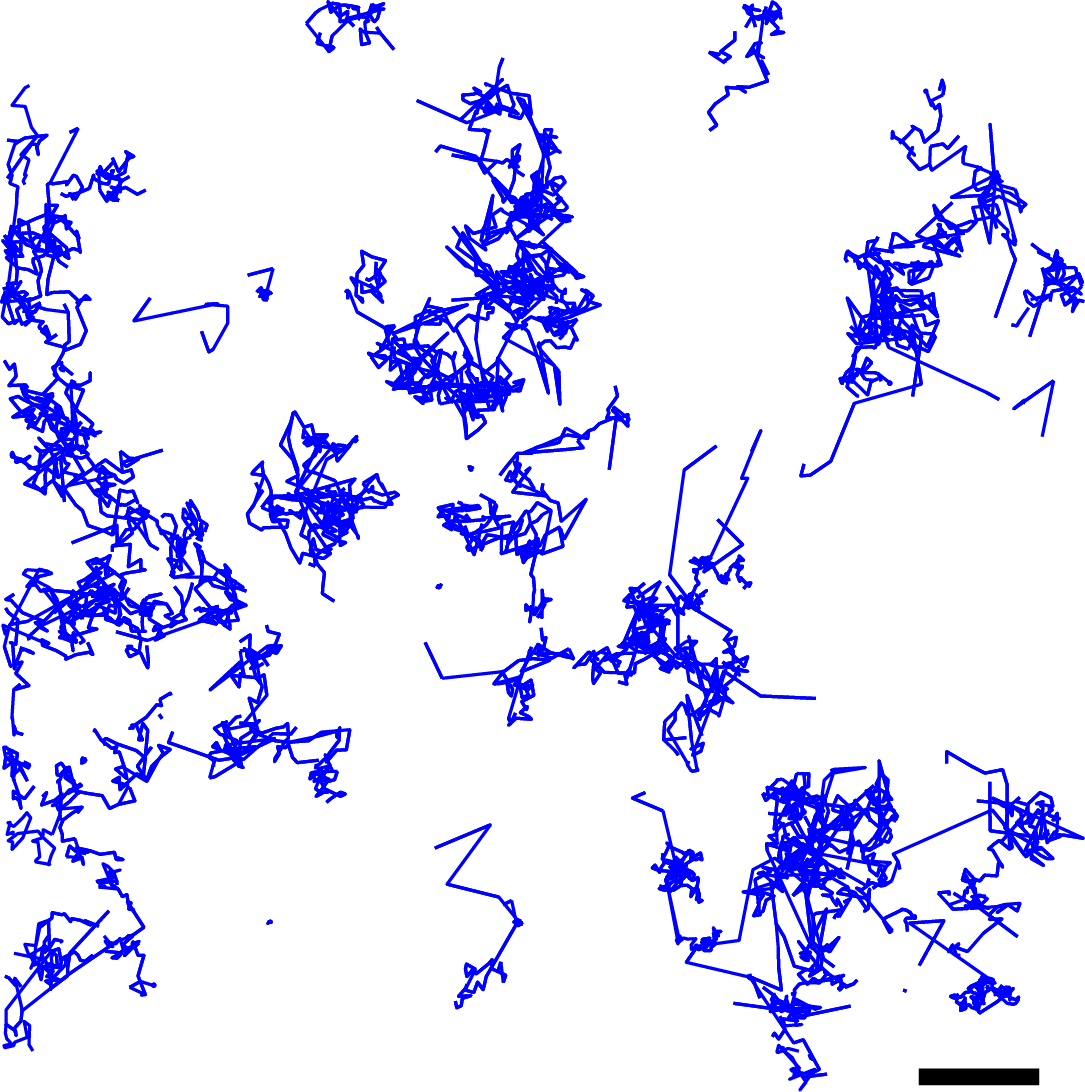}}
\vspace*{.05in}
\caption{\label{fig:tracks} 
Single particle tracking of membrane-targeting C2 domains. 
Individual trajectories are collected during 10 seconds in a 
$50 \times 50 \ \mu \textrm{m}^2$ window.  
Scale bar 5~$\mu \mathrm{m}$.}
\end{figure}

\begin{figure}
\centerline{\includegraphics[width=8 cm]{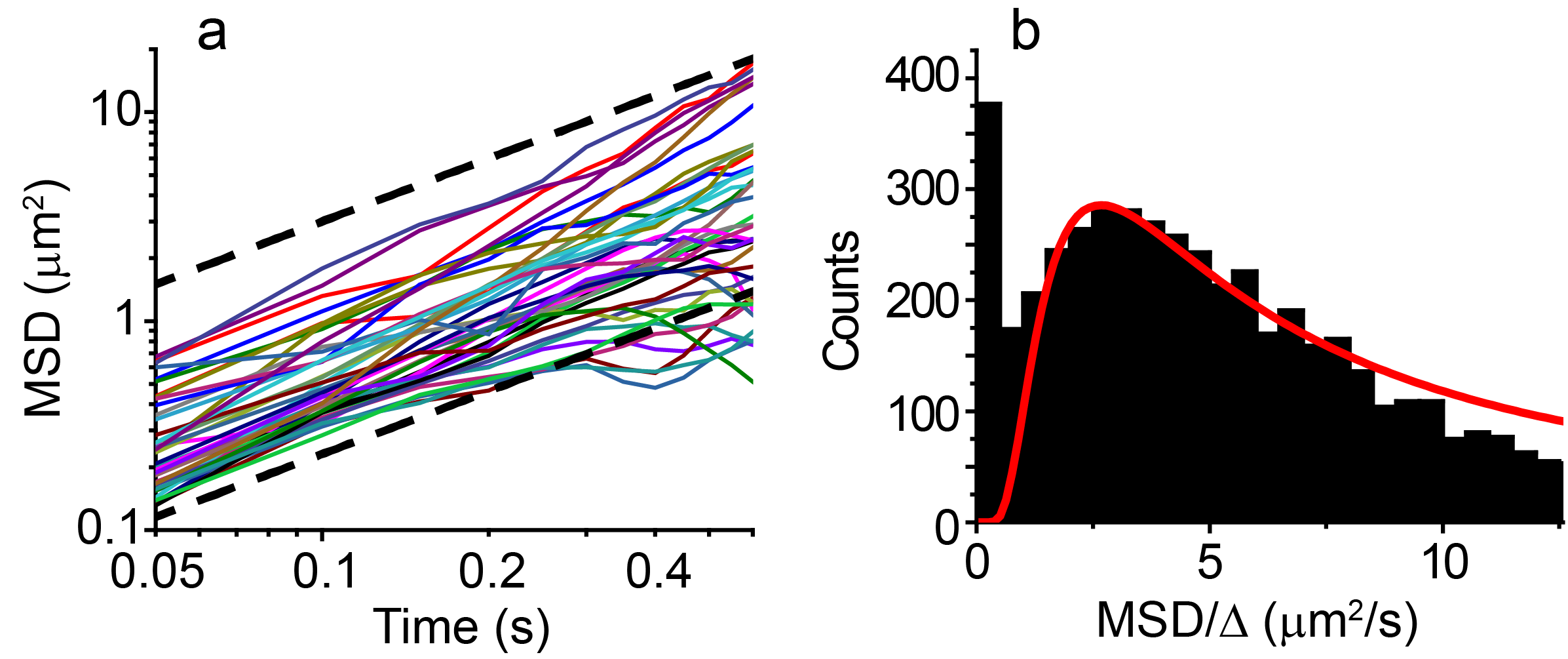}}
\caption{\label{fig:D_data} 
Scattering of the estimated diffusion coefficients of individual trajectories. 
(a) The time-averaged MSD of individual trajectories, $\overline{\delta^2(\Delta)}$, 
displays large fluctuations indicating that the MSD does not self average. 
40 randomly selected trajectories are presented in a log-log plot. 
The dashed lines are guides to the eye with $\overline{\delta^2(\Delta)}\sim \Delta$. 
The experimental time of all trajectories is 1.3 s.
(b) Distribution of the MSD slopes for C2 domains. 
The apparent diffusion coefficient can be calculated from the MSD slope, 
$\textrm{MSD}/\Delta=4D$. The thick red line 
shows the prediction by a bulk-mediated diffusion model as explained in the text.}    
\end{figure}

We tracked the motion of membrane-targeting C2A domains \cite{campagnola2015superdiffusion}, 
fluorescently labeled with Atto-565, 
on a supported lipid bilayer. Imaging was done in a home-built total internal reflection (TIRF) microscope 
under continuous illumination at 20 frames/s. 
Single-particle tracking is performed under conditions where the surface density is low 
enough to enable connections of long jumps while avoiding misconnections due to crossover between trajectories.
Figure \ref{fig:tracks} shows an example of single-molecule trajectories during 10 seconds.
As a first step, we characterize the diffusion 
by analyzing the MSD as a function of lag time. 
For each individual trajectory, the time-averaged MSD (TA-MSD) is calculated as 
\begin{equation}
\overline{\delta^2(\Delta)} = \frac{1}{t-\Delta} \int_{0}^{t-\Delta} [\textbf{r}(\tau+\Delta)-\textbf{r}(\tau)]^2 \rm d\tau,
\label{taMSD}
\end{equation}
where $\Delta$ is the lag time, $t$ the experimental time, and $\textbf{r}$ the two-dimensional 
position of a particle. 
Across the manuscript we employ brackets 
to denote the ensemble average of an observable $\langle \cdot \rangle$ 
and an overline to denote time averages $\overline \cdot$. 
Figure \ref{fig:D_data}(a) shows that, within experimental error, the TA-MSD of individual trajectories 
is linear in lag-time, resembling pure Brownian motion. 
In two dimensions, the MSD of a Brownian particle is determined by the 
diffusion coefficient $D$ via the relation $\overline{\delta^2(\Delta)} =4D\Delta$,  
but Fig. \ref{fig:D_data}(a) shows that the TA-MSD exhibits broad fluctuations. 
In ergodic systems, the time-averaged MSD converges to the ensemble average. 
In other words the time-averaged MSD can be used to consistently estimate the diffusion 
coefficient of a molecule. However, the large scattering seen Fig. \ref{fig:D_data}(a) 
indicates the time-averaged diffusion coefficient of 
individual molecules is a random variable, with no apparent convergence.
This observation suggests weak ergodicity is broken in the sense that time and ensemble averages 
do not converge to the same values \cite{he2008random}. 

Given that the TA-MSD is linear in lag time, one is tempted to find the diffusion 
coefficient of individual molecules from linear regression of the MSD trace. 
Figure \ref{fig:D_data}(b) shows the distribution of the slope of the TA-MSD, i.e. 
$\overline{\delta^2}/\Delta$, obtained from 5,187 trajectories. The distribution shows two different populations. 
A peak with very low diffusivities is apparent 
(sample mean $\langle \overline{\delta^2}/\Delta \rangle = 0.006 \ \mu$m$^2$/s). This population 
has a narrow distribution and it is attributed to particles that are immobilized and do not exhibit any motion.
A second population with high diffusivities has the characteristic large variations 
noted in Fig. \ref{fig:D_data}(a), with a 
mode at 2.7$\pm$0.1 $\mu$m$^2$/s but a sample mean 
$\langle \overline{\delta^2}/\Delta \rangle = 7.3 \ \mu \textrm{m}^2/\textrm{s}$.
When particles perform long jumps, a trajectory can be truncated and 
traces with higher diffusivities are lost. It is thus expected that 
the true distribution of MSDs is even 
broader because experimental tracking is biased towards lower diffusivities.

\subsection*{Stationarity and dependence on experimental time}
\begin{figure}
\centerline{\includegraphics[width=8 cm]{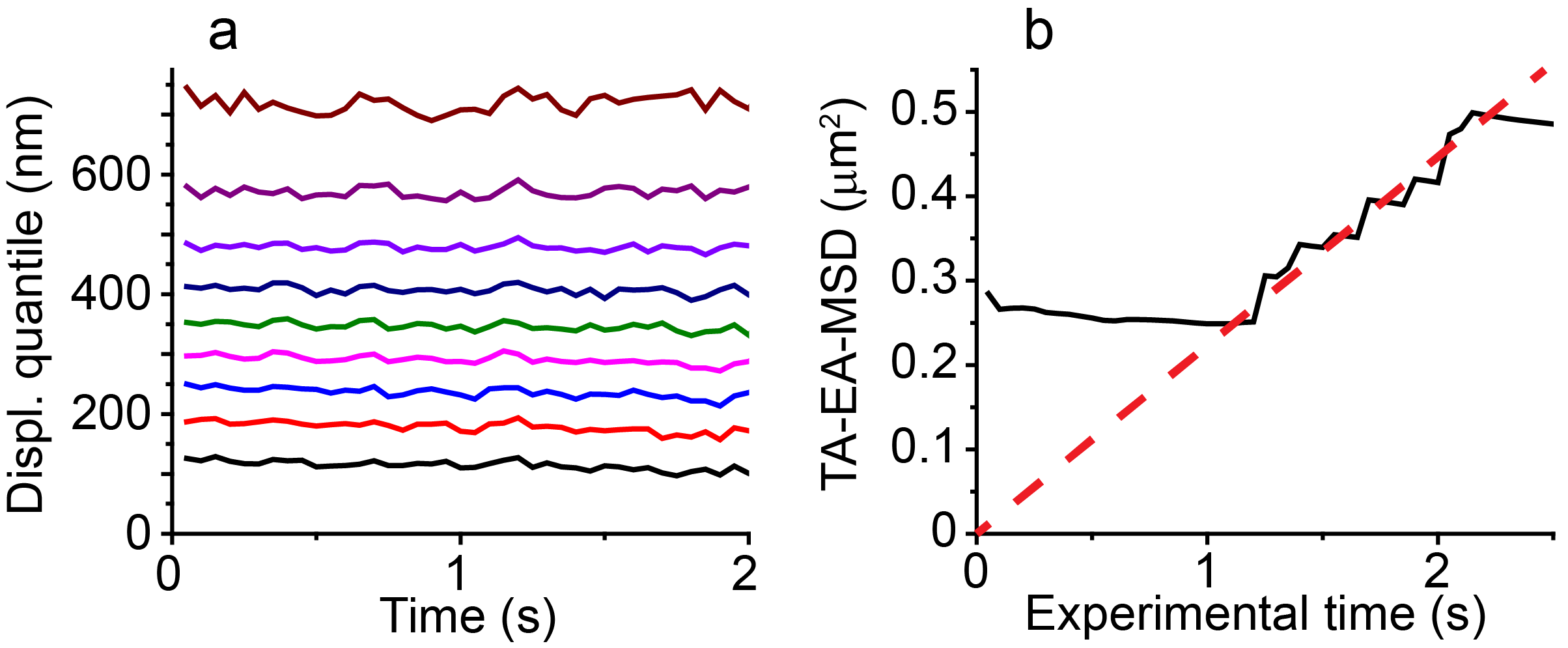}}
\caption{\label{fig:aging} 
Temporal properties of the MSD of membrane-targeting C2 domains. 
(a) 10-quantile lines of 50-ms increments of the C2 domain trajectories. 
Nine lines are shown for the fractions $k=0.1$, 0.2, ..., 0.9, 
indicating the values that divide the increments into 10 equally-populated subsets, 
each subset comprising a 10$^\textrm{th}$ of the data points.
The quantile lines are parallel, indicating the increments are stationary.
(b) Time-averaged ensemble-averaged MSD (TA-EA-MSD) as a function of observation time 
for a 50-ms lag time. 
In order to compute the TA-EA MSD, all the displacements of all trajectories up to 
time $t$ are averaged. Discrete jumps are observed, which increase the MSD
with experimental time.}
\end{figure}

It is important to establish whether the diffusion process evolves with time. Further, ergodicity 
is defined only for stationarity processes and thus we test whether the non-ergodic motion 
is rooted in the increments not being stationary. One way to check stationarity of the increments is to compute the  
quantiles as a function of time. If the quantile lines are parallel then we can infer that the process
is stationary \cite{janczura2015ergodicity}. 
Figure \ref{fig:aging}(a) shows the 10-quantile lines of the increments for lag times of 50 ms. 
The quantile lines appear to be parallel, suggesting the distribution of increments does 
not change over time. Therefore we can conclude that the process is stationary.

\begin{figure*}
\centering
\includegraphics[width=0.8 \textwidth]{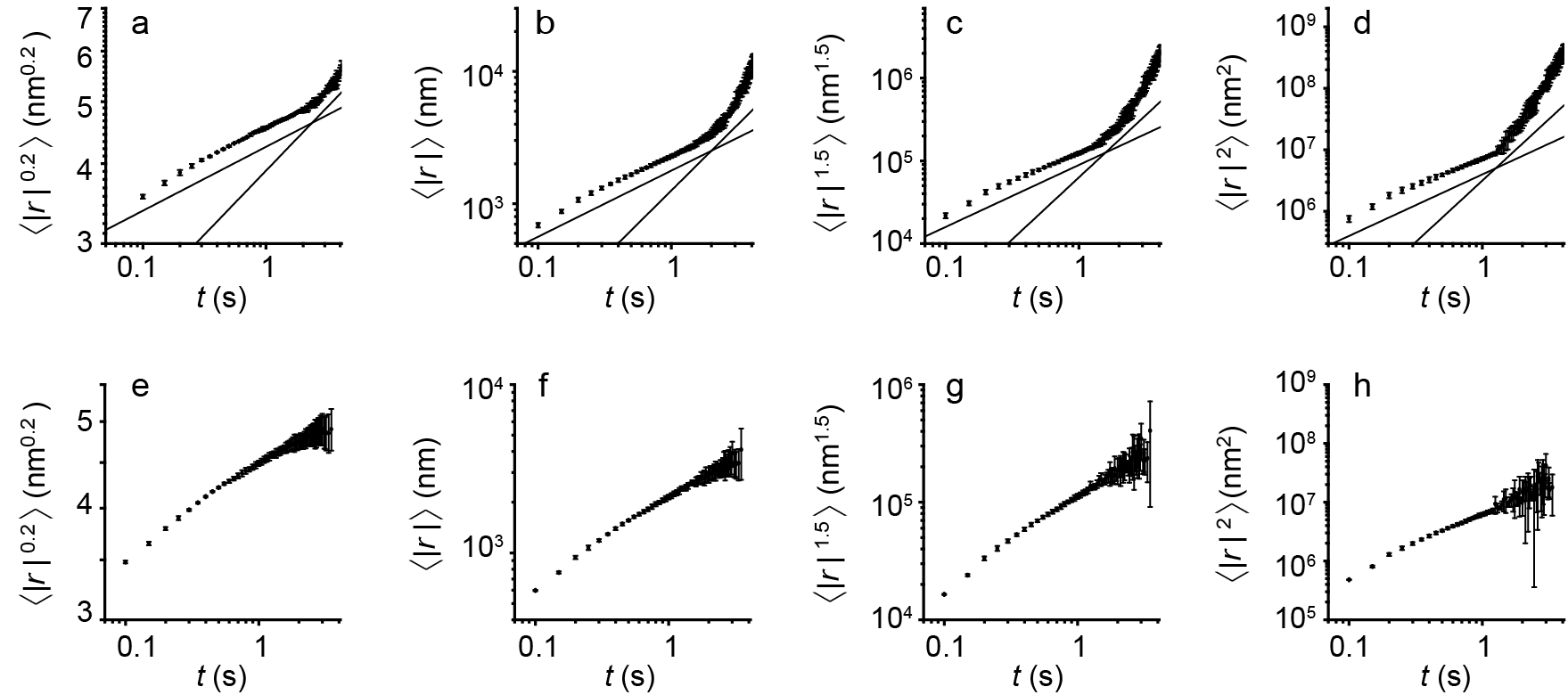}
\caption{\label{fig:qmoments} 
Ensemble-averaged $q^{\textrm{th}}$ moment of membrane-targeting C2 domains. (a-d) Moments are 
computed for $q=0.2, 1, 1.5, 2$. The solid lines provide guides to the eye 
to $\langle |r(t)|^q\rangle \sim t^{q/2}$ and $\langle |r(t)|^q\rangle \sim t^{q}$, i.e. 
$\nu=1/2$ (Brownian motion) and $\nu=1$ (superdiffusion). (e-h) The same fractional moments 
are computed when the 3\% longest displacements are excluded from the data analysis.}
\end{figure*}   
 
Even though the increments are stationary, the statistics of the diffusion process depends on experimental time. 
This effect is observed in the average of the time-averaged MSD, i.e., the time- and ensemble-averaged MSD 
(TA-EA-MSD, $\langle\overline{\delta^2}\rangle$). The TA-EA-MSD is simply the cumulative moving average of 
the square displacements, over different trajectories and for all times up to the experimental time. 
Figure \ref{fig:aging}(b) shows the TA-EA-MSD for $\Delta=50$ ms as a function of experimental time 
measured for 3,130 trajectories. The 
MSD does not appear to converge to any given value; 
instead it exhibits random jumps, so that it experiences an overall 
increase with experimental time. 
In ergodic systems, the TA-EA-MSD exhibits fluctuations around the mean, which become smaller as the available 
experimental time becomes longer due to better statistics. That type of noise is different from the 
behavior observed here because ergodicity would warrant the TA-EA-MSD converges to a finite value. 
The observed MSD increase is not monotonic and 
it decreases smoothly between jumps. Nevertheless, the rate of decrease of the MSD is much smaller 
than the average rate of increase due to the discrete jumps and thus, in probability, the MSD increases with time. 
As a consequence, if the ensemble-averaged MSD were employed to estimate a diffusion coefficient, then the coefficient 
would not be constant, but it would increase with experimental time.

\subsection*{Strong anomalous diffusion}

So far, we have characterized the dynamics of molecules using the MSD and observed that the 
TA-MSD $\overline{\delta^2}$ does not converge to the ensemble-averaged 
MSD $\langle r^2 (t)\rangle$. However, one may desire to 
characterize the motion beyond the second moment. In particular, the fractional moments 
$\langle |r(t)|^q\rangle$ with $q>0$ provide useful insight. For Brownian motion as well as 
many anomalous diffusion processes $\langle |r(t)|^q\rangle \sim t^{q\nu}$.  As long as $\nu$ is 
a constant, all moments are described by a scaling exponent linear in the order $q$
and the process is scale invariant such that the propagator at different times 
is $P(x,t)=t^{-\nu}f(x/t^{\nu})$ \cite{castiglione1999strong}. For example, in Brownian motion 
$\nu=1/2$ and $f(\cdot)$ is a Gaussian function.  

\begin{figure}
\centerline{\includegraphics[width=5 cm]{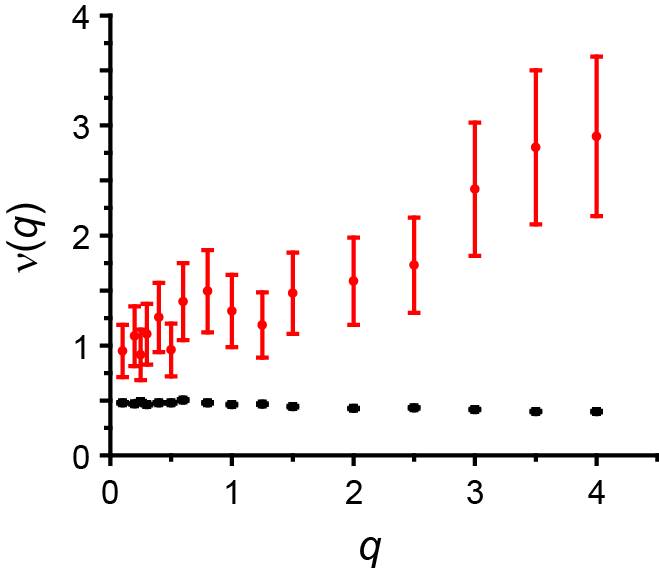}}
\caption{\label{fig:strongAD} 
The scaling exponent $q\nu(q)$ exhibits piecewise behavior at long times. At short times (lower black squares) 
$\nu(q)\approx0.5$, but at long times (upper red circles) the behavior is very different 
and $\nu(q)$ is not constant. Instead $\nu(q)$ increases with the order $q$ when $q>1$.}
\end{figure}

The process is said to exhibit strong anomalous diffusion 
when $\nu$ is not constant \cite{castiglione1999strong,metzler2014review}, 
\begin{equation}
\langle |r(t)|^q\rangle \sim t^{q\nu(q)}. 
\label{strongAnomalous}
\end{equation}
Strong anomalous diffusion has been shown 
theoretically and via numerical simulations in a variety of systems including 
the motion of tracer particles in a running sandpile model \cite{carreras1999anomalous},
the occupation times of renewal processes \cite{godreche2001statistics}, 
and flow fields \cite{castiglione1999strong} 
among others \cite{andersen2000simple,bacry2001multifractal,artuso2003anomalous}. 
In these processes, a piecewise linear scaling is found for $q\nu(q)$.
Experimental observation of strong anomalous diffusion has remained rather elusive. 
To the best of our knowledge, so far it has only been observed in the superdiffusive transport 
of polymer particles inside living cancer cells \cite{gal2010experimental}. 
Figure \ref{fig:qmoments}(a-d) shows ensemble-averaged moments of the 
two-dimensional displacements of C2 domains, which are computed by averaging over 
all available trajectories $\langle |\textbf{r}(t)-\textbf{r}(0)|^q\rangle$.
Two regimes are visible in all the moments. At short times, the fractional moments exhibit the behavior 
expected for Brownian motion, $\langle |r(t)|^q\rangle \sim t^{q/2}$, but at long times the moments ``misbehave''. 
Two solid lines are shown in each panel of Fig. \ref{fig:qmoments}(a-d): 
a shallow line with $\langle |r(t)|^q\rangle \sim t^{q/2}$ 
and a steeper line with $\langle |r(t)|^q\rangle \sim t^{q}$. For short times the agreement 
with a Brownian motion model ($q\nu(q)=q/2$) is evident. 
However, this is not the case for the long-time regime. 
In this regime, as the order $q$ increases, the logarithmic 
slopes of the moments also increase. Figure \ref{fig:strongAD} shows  
$\nu(q)$ as a function of $q$ for both the short and long times. We see that 
the scaling exponent at short times does not show significant 
deviations from $q\nu(q)=q/2$ but in the long-time regime $\nu(q)$ is not constant. 
In this time regime, $q\nu(q)=q$ for the lower order moments and 
$\nu(q)>1$ for the higher orders, which indicates strong anomalous diffusion. 
In our measurements, strong anomalous diffusion is caused 
by rare long jumps, i.e., by bulk excursions. When 
the large displacements are excluded from the analysis, the fractional 
moments display normal behavior. Figure \ref{fig:qmoments}(e-h) shows 
the fractional moments when only the displacements below a 97\% cutoff 
are considered. We observe that in this case $\langle |r(t)|^q\rangle \sim t^{q/2}$,
that is, the fractional moments without the long jumps 
scale with time as expected from Brownian motion.

\section{Theoretical Model}

\subsection*{Fluctuations in the time averages}
\begin{figure}
\centerline{\includegraphics[width=\linewidth]{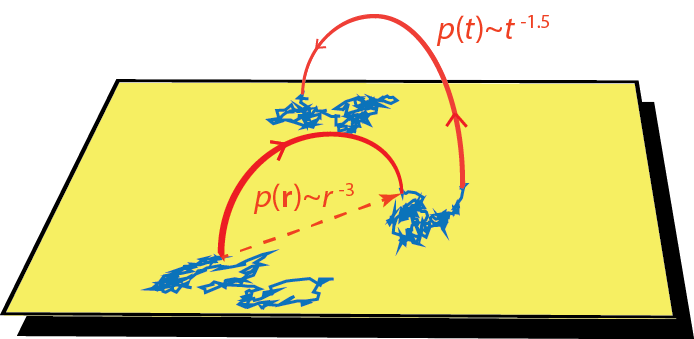}}
\vspace*{.05in}
\caption{\label{fig:sketch}Sketch of the bulk-mediated diffusion model. A molecule alternates 
between periods of 2D and 3D diffusion. The excursions into the bulk are considered as surface jumps 
with a heavy-tail distribution $p(\textbf{r})\sim r^{-3}$. In theory, the sojourn time in the bulk 
phase are asymptotically power-law distributed, $p(t)\sim t^{-1.5}$, but in practice 
jumps are observed to take place faster than the frame rate.}
\end{figure}

We have previously shown \cite{campagnola2015superdiffusion} that membrane-targeting domains 
can transiently dissociate from the lipid bilayer to perform bulk excursions. During these excursions, 
a molecule undergoes three-dimensional diffusion until it readsorbs on the surface. 
Within the bulk phase, the height $z$ 
is modeled as a one-dimensional random walk and thus the first return time distribution 
satisfies $p(t_b)\sim t_b^{-1.5}$, where the first return time $t_b$ represents 
the time the particle spends in the bulk during a single jump. 
A sketch of the model is shown in Fig.~\ref{fig:sketch}.
A simple derivation \cite{redner} leads to a one-sided L\'{e}vy distribution 
of index $1/2$, also known as a L\'{e}vy-Smirnov distribution,
\begin{equation}
p(t_b)=z_0 \left(4\pi D_b t_b^3\right)^{-1/2} \mathrm{exp}\left(-z_0^2/4D_b t_b\right), 
\label{Levy-Smirnov}
\end{equation}
where $D_b$ is the diffusion coefficient in the bulk, 
and $z_0$ is a scaling constant with units of length. 
Then, the distances on the surface covered during bulk excursions are 
two-dimensional Cauchy random variables \cite{Bychuk1995,chechkin2012,campagnola2015superdiffusion}
\begin{equation}
p(\textbf{r})=\frac{\gamma}{2\pi \left(r^2+\gamma^2\right)^{3/2}}, 
\label{Cauchy2D}
\end{equation}
where $\gamma$ is a constant with units of length. 
Interestingly, the expected values of both the first return time 
and the displacement diverge. Therefore, we expect that the time-averaged MSD is governed by extreme values. 
Namely, because the TA-MSD is determined by individual long jumps, it remains a random variable, even though 
observation times may be long.

Let us first derive the distribution of time averages from intuitive scaling arguments. 
Given that one individual long jump determines the TA-MSD of an individual trajectory, 
each TA-MSD scales as the longest displacement within the trajectory,  
\begin{equation}
\overline{\delta^2}\sim \frac{1}{t}\mathrm{max}\left \{r_i^2\right\},
\label{max}
\end{equation}
where $r_i$ are the individual measured displacements. 
From Eq.~(\ref{Cauchy2D}), we can calculate the probability density of 
squared displacements and that of TA-MSD. Defining $s=r^2$, we obtain the distribution 
$p(s)=0.5\gamma(s+\gamma^2)^{-3/2}$. 
Then, we find the distribution of $\overline{\delta^2}$ from the cumulative 
distribution function of the squared displacements $F_S(s)$. 
Namely, $F_{MSD}\left(\overline{\delta^2}\right)=\left[F_S(t \overline{\delta^2})\right]^t$ 
because the displacements are independent and identically distributed. Thus 
\begin{equation}
p\left(\overline{\delta^2}\right)\sim t^2 \left( 1 - \frac{\gamma}{\sqrt{t \overline{\delta^2}+\gamma^2}} \right)^{t-1} 
\frac{0.5 \gamma}{\left(t \overline{\delta^2}+\gamma^2\right)^{3/2}},
\label{eq}
\end{equation}
where for the sake of simplicity we 
take time $t$ as the number of time intervals, i.e. the number of measured displacements. 
In the limit of large MSDs we have $t \overline{\delta^2}\gg\gamma^2$ and Eq.~(\ref{eq}) simplifies to  
\begin{equation}
p\left(\overline{\delta^2}\right)\sim 0.5 t^{1/2}\gamma \left(\overline{\delta^2}\right)^{-3/2}.
\label{scalingTA-MSD}
\end{equation}
These simple scaling arguments yield a distribution of TA-MSD that has a power law tail with an exponent $3/2$.

Now, we follow the derivation by Froemberg and Barkai to find the whole distribution of TA-MSDs \cite{froemberg2013random}. 
In order to simplify the analysis we focus on a one-dimensional L\'{e}vy flight but the 
extension to two dimensions is straightforward.
Again the displacements are Cauchy distributed (Eq.~\ref{Cauchy2D}), albeit in one dimension, 
\begin{equation}
p(x_i)=\frac{\gamma}{\pi \left(x_i^2+\gamma^2\right)},
\label{Cauchy1D}
\end{equation}
and the square displacements $y=x^2$ are distributed according to
\begin{equation}
p(y)=\frac{\gamma}{\pi \left(y+\gamma^2\right)\sqrt{y}}\sim y^{-3/2}.
\label{PDFy}
\end{equation}
where $y\ge 0$. 
The displacements after time $\Delta$  
are  $x_\Delta=\sum_i^\Delta x_i$, with a characteristic function 
$\phi(k)=\textrm{exp}(-\gamma \Delta |x|)$. Thus 
$p(x_\Delta)=\gamma t [\pi (x_\Delta^2+\gamma^2 t^2)]^{-1}$, also a Cauchy distribution 
with a scale parameter $\gamma \Delta$. This behavior is due to the fact that the Cauchy distribution 
is stable, namely a symmetric L\'{e}vy stable distribution of index 1. 
Therefore, we can solve for $\Delta=1$ and our results 
are still valid for any lag time after rescaling $\gamma \to \gamma\Delta$.

As in Eq.~\ref{taMSD}, the TA-MSD at a lag time $\Delta$, measured over a time $t$, is \cite{froemberg2013random}
\begin{eqnarray}
\overline{\delta^2(\Delta)} & = &\frac{1}{t-\Delta} \sum_{i=1}^{t-\Delta} \left(x_{i+\Delta}-x_i\right)^2 \nonumber \\
 & {\buildrel d \over \approx} & \frac{\Delta}{t} \sum_{i=1}^{t} x_i^2,
\label{eq2}
\end{eqnarray}
where the approximation holds for $t\gg 1$. We next define the variable $\zeta=t \overline{\delta^2}/\Delta  {\buildrel d \over \approx} 
\sum_{i=1}^t x_i^2$, which is a sum of independent and identically distributed (i.i.d.) random variables $y_i$. 
Given that the variance of $y_i$ diverges, the central limit theorem breaks down and 
the distribution of $\zeta$ is found using the generalized central limit theorem \cite{feller2008introduction}. 
The Laplace transform of the distribution of $y=x^2$ (see Eq.~\ref{PDFy}) is  
\begin{eqnarray}
p\left(u_y\right) & = &\exp\left(\gamma^2 u_y \right) \mathrm{erfc} \left(\gamma\sqrt{u_y} \right) \nonumber \\
 & \approx &1 - \frac{2\gamma}{\sqrt{\pi} } \sqrt{u_y} +O(u_y) \nonumber \\
 & \approx &\exp\left(- \frac{2\gamma}{\sqrt{\pi} } \sqrt{u_y} \right),
\label{LaplaceY}
\end{eqnarray}
where $\mathrm{erfc}(\cdot)$ is the complementary error function. We 
are concerned with large values of $y$ and therefore we only keep the first term in the series expansion in 
Eq.~(\ref{LaplaceY}), that is we only consider the small $u_y$ limit in Laplace domain.  The distribution 
of $\zeta$ in the large $t$ limit is found in Laplace domain
\begin{equation}
p(u_{\zeta})=\exp\left(- \frac{2\gamma t}{\sqrt{\pi} } \sqrt{u_{\zeta}} \right).
\label{LaplaceZ}
\end{equation}
The inverse Laplace transform yields
\begin{eqnarray}
p(\zeta) & = &\frac{\pi}{2\left(\gamma t\right)^2} L_{1/2,1}\left[\frac{\pi}{2\left(\gamma t\right)^2}\zeta\right] \nonumber \\
 & = &\left(\frac{1}{2\pi c^2 \zeta^3}\right)^{1/2} \exp\left(-\frac{1}{2c^2\zeta}\right),
\label{PDFzeta}
\end{eqnarray}
where $L_{1/2,1}(\xi)$ is again the L\'{e}vy-Smirnov distribution and 
we introduced the constant $c=\sqrt{(\pi/2)}/\gamma t$. 
We can then change variables to obtain the distribution of the slope of the TA-MSD.
By defining $\xi=\zeta/t=\overline{\delta^2}/\Delta$, Eq.~\ref{PDFzeta} 
simplifies to  
\begin{equation} 
p(\xi)=\left(\frac{\gamma^2 t}{\pi^2 \xi^3} \right)^{1/2} \exp \left(-\frac{\gamma^2 t}{\pi \xi} \right). 
\label{PDFxi}
\end{equation}
Thus we find that the probability density function of the TA-MSD is a L\'{e}vy-Smirnov distribution 
with scale parameter $2 \gamma^2 t/\pi$. Recall that we derived this distribution for $t$ and $\Delta$ in number 
of frames.
In agreement with the scaling arguments discussed 
above, $p(\xi)\sim \xi^{-3/2}$. Importantly, the moments of this distribution diverge, causing large variations in the 
TA-MSD measurements as observed in Fig. \ref{fig:D_data}(a).

\subsection*{Fractional moments}
Our L\'{e}vy flight model involves a tail in the distribution of 
displacements that scales as $p(\textbf{r})\sim r^{-3}$ at long distances. Therefore, 
the $q^{\textrm{th}}$ moment diverges for $q \ge 1$. Explicitly,
\begin{equation}
\langle |r(t)|^q\rangle \sim \bigg\{
  \begin{tabular}{cc}
  $t^q$, & $q<1$  \\
  $\infty$, & $q \ge 1.$   
  \end{tabular}
\label{momentsCauchy}
\end{equation}
Of course this result is not realistic. The 
problem arises in the approximation that bulk excursions take place instantaneously. While 
the approximation is good within our experimental times, it does not hold for very long jumps, 
thus placing a bound on the higher order moments. Precise mathematical analysis that includes 
the time incurred by a bulk mediated jump would lead to the correct 
higher order moments \cite{metzler2014review,rebenshtok2014non}. 
However, a simple model leading to Eq.~(\ref{momentsCauchy}) yield some useful insights. In particular,
we can see that there is a critical order $q_c=1$ below which $\nu(q)=1$. Furthermore, 
for values $q<q_c$, the fractional 
moments yield superdiffusive behavior, i.e $\nu(q)>1/2$ as would be determined by Brownian motion. 
Above this critical value, 
the fractional moments increase above 1. The piecewise behavior is the fingerprint of strong anomalous 
diffusion as observed in Fig. \ref{fig:strongAD}.

\section{Numerical Simulations}
We test the predictions of our model using numerical simulations and compare them to the experimental data. 
Our simulations intend to model a process where molecules diffuse on a two dimensional surface 
and undergo dissociation into the bulk phase. Dissociation is considered as a Poisson process 
and the particle goes through 3D diffusion in the bulk until it finds its way back to the surface.  
5000 realizations were 
simulated off-lattice where tracers perform a random walk with Gaussian displacements in two dimensions, and 
at random times the tracer performs bulk excursions \cite{campagnola2015superdiffusion}. 
The sojourn times within the surface are exponentially distributed 
with a mean of 10 and the surface diffusion coefficient is taken to be $D_s=0.5$. 
The return times from bulk excursions are drawn from a distribution 
$\psi(t_b)= (4\pi t_b^3)^{-1/2} \mathrm{exp}(-1/4 t_b)$ (see Eq.~\ref{Levy-Smirnov}). 
Then the jump distances are Gaussian with variance $\sigma_b^2=2 t_b$.

\begin{figure}
\centerline{\includegraphics[width=8 cm]{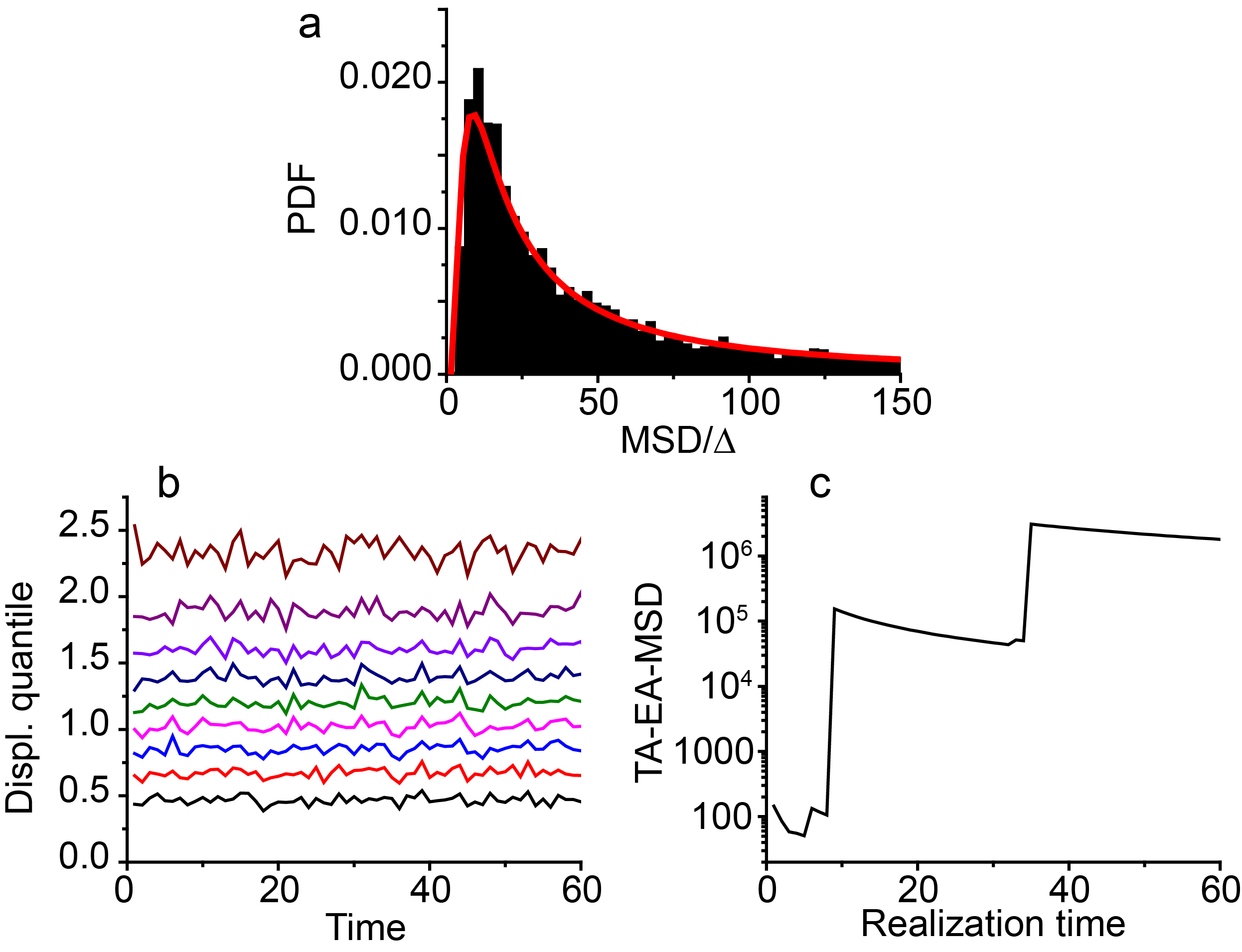}}
\caption{\label{fig:simulations} 
Model where a tracer diffuses on a plane and is allowed to dissociate to performed bulk excursions 
until readsorbing to the surface.
(a) Probability density function of the distribution of TA-MSD slopes obtained from 5,000 realizations,
where each realization includes 500 displacements. The predicted Levy distribution for the MSD 
is also shown as a solid red line.
(b) 10-quantiles of the increments of the realizations. The quantile lines are parallel, indicating that 
the process is stationary as expected \cite{janczura2015ergodicity}.
(c) The ensemble average of the TA-MSD exhibits jump discontinuities increasing the MSD when the 
realization time increases. }    
\end{figure}

Similar to the experiments on lipid bilayers, the TA-MSD of the simulations 
exhibit a broad scattering. Figure \ref{fig:simulations}(a) shows the distribution of TA-MSDs for the 
individual realizations. Overlaid on this distribution in Fig. \ref{fig:simulations}(a), 
equation \ref{PDFxi} shows good agreement with the MSD distribution. 

In our derivation of the distribution of the TA-MSDs, we have employed the Cauchy distribution (Eq.~\ref{Cauchy2D}) 
for the displacements. This equation ignores the Gaussian component in the distribution 
of displacements that arises due to the diffusive motion on the surface \cite{campagnola2015superdiffusion}.  
As seen in Fig. \ref{fig:simulations}(a), this approximation does not alter the distribution, at least in 
the long measurement time limit. The reason is that, as discussed above, the MSD is governed by the large 
displacements, i.e., the tail of the distributions. Further, we would achieve the same results (Eq. \ref{PDFxi}) 
if we only consider the power law tail of the propagator, $p(\textbf{r})\sim |r|^{-3}$ and find 
the Laplace transform using the Tauberian theorem \cite{feller2008introduction,klafter2011first}.

\section{Discussion}
In a similar fashion to the numerical simulations of bulk-mediated diffusion, 
Eq.~\ref{PDFxi} is used to model the experimental results 
for membrane-targeting C2 domains (red solid line in Fig. \ref{fig:D_data}(b)). Even though the agreement between 
our bulk-mediated diffusion model and the experimental results is satisfactory, the tail in the MSD distribution 
of C2 domains decreases faster than predicted by the model. This effect is caused by an artificial 
truncation of the distribution of displacements caused by the tracking algorithm \cite{campagnola2015superdiffusion}. 
Namely, if a particle experiences a very long jump, 
it is not possible to make frame-to-frame connections with reasonable confidence and 
thus trajectories are cropped missing the long displacements and in turn the large diffusivites.  

The increments in the motion of C2 domains on a lipid bilayer are shown to be stationary but the MSD depends on the 
experimental time (Fig. \ref{fig:aging}). This behavior is also observed in our numerical simulations. 
The increments in the simulations are stationary (Fig. \ref{fig:simulations}(b)) 
as the displacements are simulated with the 
same time-independent stochastic process. However, the TA-EA-MSD of the numerical simulations also shows 
a strong dependence on 
realization time (Fig. \ref{fig:simulations}(c)). 
In agreement with the C2 data, the simulations MSD show discrete jumps 
in the time series. Also here, the MSD average increases in probability with realization time.  

The discontinuities in the MSD as a function of experimental time can be conceptually understood in terms of the 
same mechanism that causes weak ergodicity breaking. As discussed before, the estimated diffusivities of individual 
trajectories are governed by extreme displacements. Recall that the reason for the lack of self-averaging is  
the existence of one displacement in the trajectory that is likely 
much larger than all others and thus the MSD depends on this 
individual displacement. In the same way, at a given time $t$, a jump may occur among all the molecules such that it is much larger 
than all the displacements observed thus far. When such an event takes place the TA-EA-MSD increases sharply due to the contribution 
of one long jump. After a very large jump occurs, the relative weight of that individual 
long jump diminishes because more data points become available. 
Thus, following a jump discontinuity the MSD decreases with experimental time. The 
MSD continues to decrease until the next jump discontinuity takes place.
 
We observed that, in probability, the sample mean of the TA-MSD increases with experimental time. 
This is also observed in the theoretical distribution of the TA-MSD (Eq.~\ref{PDFxi}), 
which involves a scale parameter that explicitly depends on the experimental time.
Even though the expected value of the TA-MSD diverges we can estimate how the MSD increases with experimental time 
by evaluating other measures of central tendency, such as the theoretical mode and median. 
Both of these measures scale linearly with time, namely. 
\begin{equation} 
\textrm{mode}_{\xi}=\frac{2\gamma^2}{3\pi}t 
\label{mode}
\end{equation}
and
\begin{equation} 
\textrm{median}_{\xi}=\frac{\gamma^2}{\pi \left[\textrm{erfc}^{-1}(1/2)\right]}t 
\label{median}
\end{equation}
where $\textrm{erfc}^{-1}(\cdot)$ is the inverse complementary error function. Thus we expect the 
average of the TA-MSD to increase in probability linearly 
with experimental time as observed in Fig. \ref{fig:aging}(b).

In this manuscript we employed the fractional moments and showed that the system exhibits strong 
anomalous diffusion. The fractional moments are only rarely used in the diffusion 
literature. Nevertheless, these moments can be very useful in the analysis of 
bulk mediated diffusion. When the trajectories are modeled as L\'{e}vy flights 
the theoretical MSD diverges and its use in the analysis of motion 
challenging. This is not the case for fractional moments with $q<1$, 
where the theoretical moment is finite and no discontinuities are observed. 
Thus low order moments become a useful tool to study phenomena such as 
superdiffusion.

Our data shows that the common practice of finding diffusivities from time averaged MSD in membranes should be 
approached with care. We find that weak ergodicity is broken and as a consequence the 
MSD of individual trajectories are random variables even in the long time limit. In other words, the MSD from 
individual trajectories are not reproducible. Furthermore, the ensemble mean of the time-averages is not a reliable measure 
because it depends on experimental time. Careful analysis indicates that as the available measurement time becomes longer, 
the apparent diffusion coefficient increases. In order to deal with this subtlety we propose that, 
when bulk excursions are evident in the data, parameters are extracted from the distributions instead of using 
either time or ensemble averages. We have previously shown that it is feasible to obtain 
both the surface diffusion coefficient 
and the scale parameter $\gamma$ from the distribution of displacements 
when the data sample is large enough \cite{campagnola2015superdiffusion}.

\section{Conclusions}
We have shown that bulk-mediated diffusion can be accurately modeled as a L\'{e}vy flight. 
The L\'{e}vy flight concept 
yields superdiffusive dynamics with complex strange kinetics, in particular because the time averaged MSD does not 
converge to the ensemble average. Thus the process exhibits weak ergodicity breaking. 
The time-averaged MSD of individual trajectories is governed by individual long jumps and, 
as a consequence, it remains a random variable. 
We have shown that the MSD also depends on experimental time and thus 
it does not provide a consistent estimator of the diffusion coefficient. The long time asymptotic of 
the displacement fractional moments has the signature of superdiffusive behavior both for low and high orders. 
Moreover, the L\'{e}vy flight model predicts strong anomalous diffusion, 
a phenomenon that deals with non-linear scaling exponents 
of the fractional displacement moments. We have experimentally observed 
this anomalous behavior in the motion of membrane-targeting domains on supported lipid bilayers 
using single-particle tracking. Furthermore, given the broad 
applicability of bulk mediated diffusion, we foresee these anomalies can be observed in many complex systems.

\section{Methods} 

\subsection*{Preparation of supported lipid bilayers}
Lipid bilayers were prepared as described elsewhere \cite{campagnola2015superdiffusion}. 
In brief, chloroform-suspended 18:1 ($\Delta 9$-Cis) PC (DOPC) and 18:1 PS (DOPS) were mixed at a ratio of 3:1. 
The phospholipid mixture was vacuum dried overnight and 
resuspended in imaging buffer (50 mM HEPES, 75 mM NaCl, 1 mM MgCl$_{2}$, 2 mM tris(2-carboxyethyl)phosphine (TCEP), 200 $\mu$M CaCl$_{2}$) to a final concentration of 3 mM followed by probe sonication 
to form sonicated unilamellar vesicles (SUVs) \cite{corbin2004grp1}. 
A solution of SUVs (1.5-mM lipid) in 0.5 M NaCl and imaging buffer was introduced into a perfusion chamber 
(CoverWell, Grace Bio-Labs model PC8R-1.0)
and incubated for one hour at 4$^{\circ}$C. 
The surface was then rinsed with imaging buffer multiple times prior to addition of protein sample. 

\subsection*{Protein expression, purification, and labeling}
An expression plasmid containing the ybbr-Synaptotagmin 7 (Syt7) C2A gene \cite{ybbr} 
was transformed into \textit{Escherichia coli} BL21-CodonPlus(DE3) competent cells. 
Cells were grown at 37$^{\circ}$C to an OD$_{600}$ of 0.6 
and then induced to express protein with 0.5 mM IPTG for 6 hours at room temperature.
The harvested cells were lysed at 18,000 lb/in$^2$ in a microfluidizer 
in a buffer containing 50 mM Tris, pH 7.5, 400 mM NaCl and centrifuged at 17,000 rpm.
The clarified lysate was loaded onto a 5-ml GSTrap FF column (GE Healthcare LifeSciences, Pittsburgh, PA) 
followed by gradient elution with 50 mM Tris, pH 8.0, 100 mM NaCl, and 10 mM glutathione. 
Fractions containing protein were pooled and diluted to reduce the salt to less than 0.1 M prior 
to loading onto a HiTrap Q HP column (GE Healthcare LifeSciences, Pittsburgh, PA) and 
eluting with a linear gradient to 1 M NaCl in 25 mM Tris, pH 8.5, 20$\%$(vol/vol) glycerol, and
0.02$\%$(wt/vol) NaN$_3$. A portion of the purified protein was subjected to thrombin 
cleavage to remove the GST tag and then separated using a Superdex 200 gel filtration column 
(GE Healthcare LifeSciences, Pittsburgh, PA) equilibrated in 50 mM Tris, pH 7.5 and 100 mM NaCl. 

20 mM CoASH (New England Biolabs, Ipswich, MA) in 400 mM Tris, pH 7.5 was mixed with 20 mM 
Atto-565 maleimide (ATTO-TEC, Siegen, Germany) in dimethylformamide and incubated at 30$^{\circ}$C overnight
to form Atto-565 CoA, then diluted 10 fold with 5 mM DTT, 10 mM Tris pH 7.5 to quench the reaction. 
ybbr-Syt7 C2A was labeled with the Atto-565 via SFP synthase (4$\prime$-phosphopantetheinyl transferase). 
Samples were dialyzed against 1 L of 50 mM HEPES, pH 7.0, 75 mM NaCl, 4 mM MgCl$_{2}$ and
5$\%$ glycerol overnight at 4$^{\circ}$C, and then concentrated to 10 $\mu$M.

\subsection*{Imaging and single-particle tracking}
Proteins were added to the imaging buffer to a final concentration of 75 pM. Then, 
the perfusion chamber was filled with the solution. The perfusion chambers were 
9 mm in diameter and 0.9-mm deep, holding a volume of $\approx 60\ \mu$l. 
Imaging was performed without replacing the solution, so that 
there was always protein present in the bulk solution and the surface concentration could reach a steady state.

All images were acquired using an objective-type total internal reflection fluorescence
microscope (TIRFM) as described previously \cite{weigel2011PNAS,weigel2013pnas}. 
A 561 nm laser line was used as excitation source. 
A back-illuminated electron-multiplied charge coupled device (EMCCD)
camera (Andor iXon DU-888) liquid-cooled to -85$^{\circ}$C, with an electronic gain of 300 was used.
In order to maintain constant focus during the whole imaging time we employed an autofocus system 
(CRISP, Applied Scientific Instrumentation, Eugene, OR) in combination with a piezoelectric stage 
(Z-100, Mad City Labs, Madison, WI). Videos were acquired at a frame rate of 20 frames/s
using Andor IQ 2.3 software and saved as 16-bit tiff files. 
The images were filtered using a Gaussian kernel with a standard deviation of 1.0 pixel in ImageJ.
Single-particle tracking of Atto-C2 was performed in MATLAB 
using the u-track algorithm developed by Jaqaman et al. \cite{jaqaman2008U-track}.

\begin{acknowledgments}
We thank Professor Eli Barkai for his useful suggestions and discussions and 
Professor Jeff Knight for kindly supplying the C2A plasmid.
KN thanks Bryce Schroder and Sanaz Sadegh for their help with the experimental setup. 
We acknowledge the support from the National Science 
Foundation under grant 1401432 and the National Institutes of Health under grant R21AI111588.
\end{acknowledgments}

\bibliography{Distbib} 

\end{document}